\documentclass[twocolumn,pra]{revtex4}

\usepackage{graphicx}
\usepackage{dcolumn}
\usepackage{bm}
\usepackage{theorem}
\newtheorem{definition}{Definition}[section]
\newtheorem{proposition}[definition]{Proposition}

\def\squareforqed{\hbox{\rlap{$\sqcap$}$\sqcup$}}
\def\qed{\ifmmode\squareforqed\else{\unskip\nobreak\hfil
\penalty50\hskip1em\null\nobreak\hfil\squareforqed
\parfillskip=0pt\finalhyphendemerits=0\endgraf}\fi}
\def\endenv{\ifmmode\;\else{\unskip\nobreak\hfil
\penalty50\hskip1em\null\nobreak\hfil\;
\parfillskip=0pt\finalhyphendemerits=0\endgraf}\fi}
\newenvironment{proof}{\noindent \textbf{{Proof~} }}{\qed}

\newcommand{\beq}{\begin{equation}}
\newcommand{\eeq}{\end{equation}}
\newcommand{\beqa}{\begin{eqnarray}}
\newcommand{\eeqa}{\end{eqnarray}}
\newcommand{\beqar}{\begin{eqnarray*}}
\newcommand{\eeqar}{\end{eqnarray*}}

\def \ra {\rangle}

\begin{document}
\input epsf
\title{\bf \large Reliable entanglement transfer between pure quantum states}

\author{Berry Groisman}
\affiliation{Centre for Quantum Computation, DAMTP, Centre for
Mathematical Sciences, University of Cambridge, Wilberforce Road,
Cambridge CB3 0WA, United Kingdom. }


\begin{abstract}
The problem of the reliable transfer of entanglement from one pure
bipartite quantum state to another using local operations is
analyzed. It is shown that in the case of qubits the amount that can
be transferred is restricted to the difference between the
entanglement of the two states. In the presence of a catalytic state
the range of the transferrable amount broadens to a certain degree.
\end{abstract}


\maketitle

\section{Introduction}\label{intro}
One of the most important recent achievements of quantum information
theory is that of establishing the paradigm of bipartite
entanglement as an asymptotically fungible resource
\cite{proc_meth}. In the asymptotic limit there are no theoretical
restrictions on entanglement manipulation: bipartite entanglement
can be redistributed reliably (i.e., without losses) as desired. It
can be, for example, concentrated in a small number of states or
diluted into a larger number of states.

In real situations, however, we always deal with a finite number of
entangled states, and it is of practical importance to know what
kinds of manipulations of entanglement of a finite number of states
are permitted. The finite number scenario puts severe limitations on
the efficiency of entanglement manipulations. Most of the protocols
are accompanied by inevitable entanglement loss.

In this paper I address the question of whether any nontrivial
reliable manipulations of entanglement of a finite number of states
are possible. In particular, I analyze the simplest case where
parties share two pure entangled states and want to transfer a
fraction of entanglement from one of them to another.

 Let us imagine the following situation. Alice and Bob, who live
 very far from each other,
share a pure entangled state $|\psi\ra_{AB}$, where $A$ and $B$ are
Alice's and Bob's quantum particles respectively. Alice and Bob have
two friends, Alex and Barbara, who also share a pure entangled state
$|\phi\ra_{ab}$ between them. Alex lives in the same city with
Alice, so any joint task carried out by Alice and Alex can be
regarded as local. Similar rules apply to Bob and Barbara. The
problem is formulated as follows: Is it possible to design a LOCC
protocol which will transfer an amount of entanglement $\Delta E$
from the ``donor" state $|\psi\ra_{AB}$ [thereby reducing its
entanglement to $E(\psi)-\Delta E$] to the ``acceptor" state
$|\phi\ra_{ab}$ [thereby increasing its entanglement to
$E(\phi)+\Delta E$]? And if yes, what are the conditions for such a
transformation? (Note that both $|\psi\ra$ and $|\phi\ra$ are
required to remain pure.)

The entanglement transfer scenario described above is relevant for
many tasks in quantum information. Recently, it has been shown that
the successful implementation of some nonlocal operations, such as
nonlocal POVM measurements \cite{nlPOVM}, requires entangled states
that possess a particular (nonmaximal) amount of entanglement; a
``catalyst" state, needed to make some entanglement transformations
possible, must be nonmaximally entangled \cite{catalysis}. In the
two above examples, if we are given the state more entangled than
required, then we obviously will have to reduce it. It is always
possible to reduce the amount of entanglement by losing part of it.
However, as entanglement is an expensive resource, we might prefer
to  transfer the redundant part to another system for future use. An
additional example is the entangling capacity of nonlocal
Hamiltonians and nonlocal unitaries. It was shown that the maximal
rate of entanglement creation is achieved when a nonlocal
Hamiltonian or a unitary acts on qubits that are partially entangled
\cite{nlham,lhl}. Thus, in order to maintain the maximal rate of
entanglement production, one would like to be able to ``transfer"
the generated gain in entanglement (``suplus value") to a different
system after each application in order to keep the target state in
its optimal form.

Clearly, there is a situation when the entanglement transfer is
possible. Indeed, Alice and Bob can locally swap the states of $A,a$ and
$B,b$, respectively, 
thereby transforming the total initial state
$|\psi\ra_{AB}\otimes|\phi\ra_{ab}$ into
$|\phi\ra_{AB}\otimes|\psi\ra_{ab}$, and transferring the amount of
entanglement $\Delta E=E(\psi)-E(\phi)$. This trivial protocol is
not really helpful, though, because it restricts the state
$|\phi\ra_{ab}$ to that which we want to obtain in the first place.
A nontrivial and interesting situation occurs when the desired state
is not possessed initially either by Alice and Bob or by Alex and
Barbara, and when $\Delta E$ is not determined by the initial
states.

On the other hand, it is clear that there are situations when
entanglement transfer is impossible. For example, let us assume that
both donor and acceptor states possess the same amount of
entanglement equal to $0.5$ ebit. If our team was able to transfer
all the amount of entanglement from the donor state to the acceptor
state (thus
 doubling the entanglement of the latter), then it would essentially mean that they
 reliably implemented the entanglement concentration in the two-copy
 scenario. Imagine that there are
 $n\gg1$ such pairs of states, and the above hypothetic protocol is implemented on each pair of
 states separately. $2n$ nonmaximally entangled states will be
 concentrated into $n$ maximally entangled states. Such a procedure
 would not only achieve the result of the collective entanglement
 concentration method \cite{proc_meth} by acting on the states
 individually, but would even outperform it as no losses, even
 sublinear, will take place. Although it might be possible to use
 such {\it reductio ad absurdum} arguments based on asymptotic
 scenarios to deduce that certain single-copy transformations are
 impossible, I will not base my argument on the asymptotic case
 at all. Instead, I will use only results and theorems for
 a single copy, making the analysis logically self-sufficient. I believe that this approach will
 give interesting and fundamental insight into the nature of entanglement of the final number of
 states.

 The structure of the article is as follows. In Sec.
 \ref{product_acceptor} I will analyze the case of a disentangled acceptor
 state for the quantum system
 of any finite dimensionality. In Sec. \ref{entangled_acceptor} the
 case of an entangled acceptor for qubits is analyzed.
In Sec. \ref{catalysis} the possibility of catalytic transformation
is taken into account. Sec. \ref{prob_transfer} demonstrates
 how all restrictions might be relaxed if probabilistic
transformation is allowed. Finally, Sec. \ref{asymptotic} shows
 that all restrictions are overcome in the
 asymptotic limit.

\section{Entanglement transfer to a direct product state}\label{product_acceptor}
Let us start with considering the case when $|\phi\ra_{ab}$ is a
direct product, i.e., $E(\phi_{ab})=0$. The results of this section
can be applied to quantum systems of any finite dimensionality.
\begin{proposition}
  \label{prop: piece_ent}
  Given a single copy of a bipartite pure entangled state $|\psi\ra$, it is
  impossible to transfer part of the amount of entanglement
  possessed by $|\psi\ra$ 
   to different quantum
  systems, which are initially disentangled, by means of LOCC without
  changing the Schmidt number of $|\psi\ra$.
 \end{proposition}
 \begin{proof}
 As a consequence of the {\it majorization condition}
 \cite{nielsen}, the Schmidt number of a quantum state cannot be
 increased by LOCC. The hypothetical transformation under question
 leads to the inevitable increase of the Schmidt number, and therefore is
 forbidden. Indeed, before the transformation, the total Schmidt
 number is equal to the Schmidt number of $|\psi\ra$. If the Schmidt number of $|\psi\ra$ does not change, then
 after the transformation the total Schmidt number equals the Schmidt number of
 $|\psi\ra$ times the Schmidt number of $|\phi\ra$ (the state which the entanglement was transferred to).
\end{proof}

{\it Corollary.} For two-qubit and two-qutrit states the
entanglement can only be transferred in full because a two-qubit
entangled state can only have the Schmidt number 2, while the next
number below is 1 for product states. For qutrit states the maximal
Schmidt number of 3 can also be reduced only to 1 (not to 2). The
task can be trivially accomplished simply by two local SWAP
operations.

The corollary and Proposition \ref{prop: piece_ent} also apply to
situations when we do allow some amount of entanglement to be lost
during the transfer. The results of this section are consistent with
the approach taken in the broadcasting of entanglement
\cite{ent_broadcast} (see a more detailed discussion in Sec.
\ref{concl}). Indeed, here it has been shown that the entanglement
of a single pure state cannot be split into two less-entangled pure
states. The only open possibility is that the states involved are
mixed - exactly the case that was analyzed in Ref.
\cite{ent_broadcast}.

Although the results of the next section are more general and
include direct product acceptor states as a special case,
Proposition \ref{prop: piece_ent} stands as an important result on
its own. The arguments used in the proof are simpler than those in
Sec. \ref{entangled_acceptor}. Besides, in Sec.
\ref{entangled_acceptor} we assume reliable protocols for qubits,
while Proposition \ref{prop: piece_ent} is valid for quantum systems
of any dimensionality and for a more general scenario when we do
allow entanglement losses.

\section{Entanglement transfer to an entangled state}\label{entangled_acceptor}
In this section I will analyze the general case of the entangled
acceptor state $|\phi\ra_{ab}$ for qubits. Let me write the donor
state in its Schmidt decomposition as \beq
|\psi_{\beta}\ra_{AB}=\cos \beta |\mu\ra_A |\nu\ra_B+\sin
\beta|\mu^{\perp}\ra_A |\nu^{\perp}\ra_B,\eeq where it is assumed
that all phases are absorbed by local basis states
$|\mu^{\perp}\ra_A$ and $|\nu^{\perp}\ra_B$. These phases, as well
as actual local basis states, are not important as we will be
interested below only in the values of the Schmidt coefficients.
Similarly, I can write the acceptor state (of the qubits $a$ and
$b$) as \beq |\phi_{\alpha}\ra_{ab}=\cos\alpha |\xi\ra_a
|\eta\ra_b+\sin \alpha |\xi^{\perp}\ra_a |\eta^{\perp}\ra_b.\eeq We
denote the amounts of entanglement possessed by
$|\psi_{\beta}\ra_{AB}$ and $|\phi_{\beta}\ra_{ab}$ as
$E(\psi_{\beta})$ and $E(\phi_{\alpha})$.

 Without loss of
generality, let me assume that $0\leq\alpha\leq\pi/4$,
$0\leq\beta\leq\pi/4$ and denote the (decreasingly ordered) Schmidt
coefficients of the donor state and the acceptor state by
$\{c_{\beta}^2, s_{\beta}^2\}$ and $\{c_{\alpha}^2, s_{\alpha}^2\}$,
respectively \cite{footnote_1}. The reduction of entanglement of
$|\psi_{\beta}\ra$ by $\Delta E$ corresponds to a reduction of
$\beta$ by $\Delta \beta$. Subsequently, the increase of
entanglement of $|\phi_{\alpha}\ra$ by the same amount $\Delta E$
corresponds to an increase of the angle $\alpha$ by $\Delta \alpha$.
Note that in general $\Delta \alpha\neq\Delta \beta$. Here we use
the entropy of entanglement as an entanglement measure, thus $\Delta
\alpha$ and $\Delta \beta$ are related by the formula
\beq\label{ent_balance}
H[c_{\beta}^2]+H[c_{\alpha}^2]=H[c_{\beta-\Delta\beta}^2]+H[c_{\alpha+\Delta\alpha}^2],
\eeq where $H[x]=-x \log_2 x-(1-x) \log_2 (1-x)$ is the (Shannon)
entropy of the probability distribution $\{x, 1-x\}$. 


Thus, there are three free parameters in the problem. I will fix
$\beta$ and $\Delta\beta$ and investigate which values of $\alpha$
are possible. As we want to avoid reducing the entanglement of
$|\phi_{\alpha}\ra$, the relevant range of $\alpha$ is
$0\leq\alpha\leq(\pi/4-\Delta\alpha)$. $\Delta\alpha$ enters the
problem as an implicit function of $\alpha$, which is determined by
Eq. (\ref{ent_balance}).

In order to analyze the possibility of such a transformation, we
employ the majorization condition \cite{nielsen}, which states that
the transformation is possible iff the ordered Schmidt coefficients
(of the combined four-qubit system) before the transformation
$\{\lambda_1,\lambda_2,\lambda_3,\lambda_4\}$ and after the
transformation $\{\lambda_1',\lambda_2',\lambda_3',\lambda_4'\}$
satisfy the following three inequalities:
\begin{eqnarray}
\lambda_1'\geq\lambda_1,\label{majo1}\\
\lambda_1'+\lambda_2'\geq\lambda_1+\lambda_2,\label{majo2}\\
\lambda_1'+\lambda_2'+\lambda_3'\geq\lambda_1+\lambda_2+\lambda_3,\label{majo3}
\end{eqnarray}
where the last inequality can be rewritten as
$\lambda_4'\leq\lambda_4$.

We can write the Schmidt coefficients in terms of the parameters of
the problem {\it up to the ordering of the second and the third
coefficients}, which depends on the particular values of $\alpha$,
$\beta$, $\Delta\alpha$ and $\Delta\beta$.
They are \beqa \{\lambda_i\}=\{(c_{\alpha}c_{\beta})^2,
(c_{\alpha}s_{\beta})^2, (s_{\alpha}c_{\beta})^2,
(s_{\alpha}s_{\beta})^2\},~~~\label{Schmidt_before}\\
\{\lambda_i'\}=\{(c_{\alpha+\Delta\alpha}c_{\beta-\Delta\beta})^2,
(c_{\alpha+\Delta\alpha}s_{\beta-\Delta\beta})^2,~~~~~\nonumber\\
(s_{\alpha+\Delta\alpha}c_{\beta-\Delta\beta})^2,
(s_{\alpha+\Delta\alpha}s_{\beta-\Delta\beta})^2\}.
\label{Schmidt_after}\eeqa It will be more convenient to use the
following inequalities that are equivalent to Eqs.
(\ref{majo1})-(\ref{majo3}):
\begin{eqnarray}
f_1\equiv\sqrt{\lambda_1'}-\sqrt{\lambda_1}\geq 0,\\
f_2\equiv\sqrt{\lambda_1'+\lambda_2'}-\sqrt{\lambda_1+\lambda_2}\geq 0,\\
f_3\equiv\sqrt{\lambda_4}-\sqrt{\lambda_4'}\geq 0,
\end{eqnarray}
which will present no problem since all trigonometric functions used
in Eqs. (\ref{Schmidt_before}) and (\ref{Schmidt_after}) are
positive in the relevant range of parameters.

 In the following paragraphs it
will be shown that for any value of
$\alpha\in[0,\pi/4-\Delta\alpha]$ at least one of the functions
$f_1$, $f_2$ or $f_3$ is negative, which will be sufficient for us
to conclude that the transformation is impossible. The only
exception is the point $\alpha_{\ast}=\beta-\Delta\beta$ which is
the simultaneous solution of $f_1=0$, $f_2=0$, and $f_3=0$. Only for
$\alpha=\alpha_{\ast}$ the transformation is possible.

\begin{figure}[h!]
\centering
$\begin{array}{cc}
\includegraphics[width=0.47\textwidth]{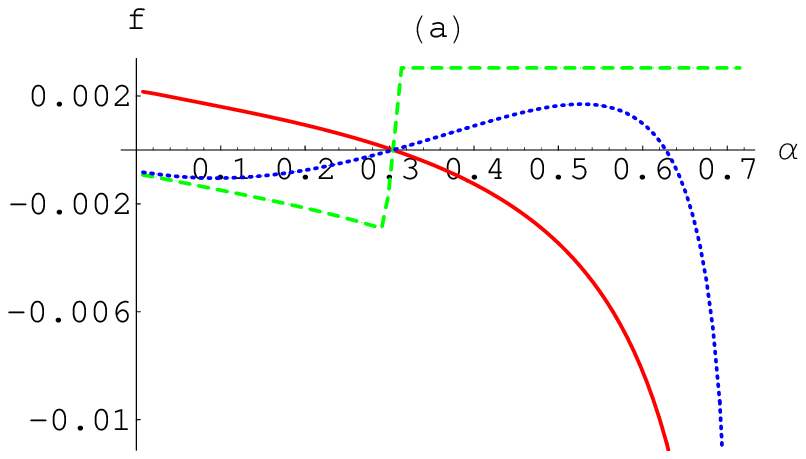}\\
\includegraphics[width=0.47\textwidth]{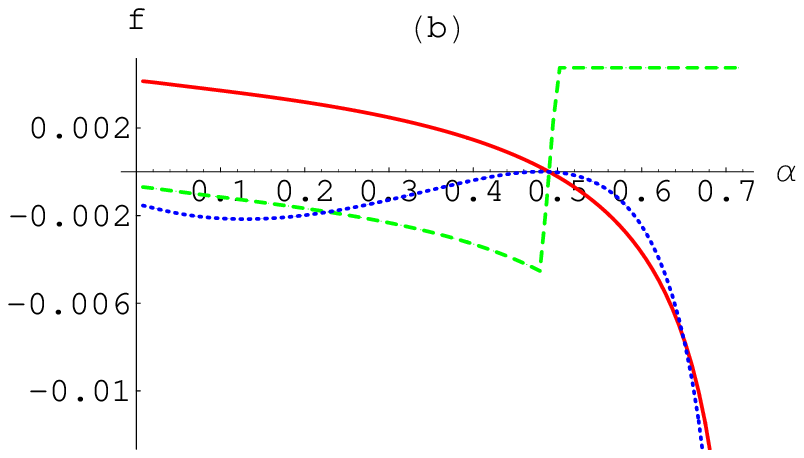}\\
\includegraphics[width=0.47\textwidth]{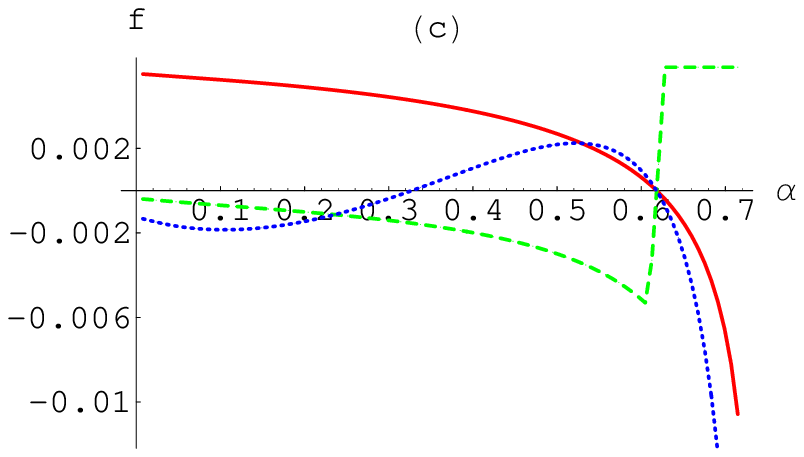}\label{fig1c}
\end{array}$
\caption{(Color online) $f_1$ (solid line), $f_2$ (dashed line), and
$f_3$($\times 10$) (dotted line) as functions of $\alpha$ for
$\Delta\beta=0.01$: (a) $\beta=\pi/10$, (b) $\beta=0.5$, (c)
$\beta=\pi/5$.
  The point $\alpha_{\ast}$ and the three different regimes of $f_2$ are clearly visible.
  Here and in all the following figures $\alpha$, $\Delta\alpha$, $\beta$ and $\Delta\beta$ are measured
  in radians.}
\label{fig1}
\end{figure}

Using Eqs. (\ref{Schmidt_before}) and (\ref{Schmidt_after}), we can
write $f_1$ and $f_3$ unambiguously without any additional
assumptions regarding the values of the parameters
\begin{eqnarray}\label{f1}
f_1=c_{\alpha+\Delta\alpha}c_{\beta-\Delta\beta}-c_{\alpha}c_{\beta}\\
f_3=s_{\alpha}s_{\beta}-s_{\alpha+\Delta\alpha}s_{\beta-\Delta\beta}.
\end{eqnarray}
There is an ambiguity regarding $f_2$ though. Depending on the
ordering of the actual Schmidt coefficients, the three following
regimes are obtained for $f_2$:
\begin{equation}\label{f2}
f_2=\left \{ \begin{array}{lcl} c_{\alpha+\Delta\alpha}-c_{\alpha}
,&& \mbox{ if
$\alpha<\beta-\Delta\alpha-\Delta\beta$}\\
c_{\beta-\Delta\beta}-c_{\alpha},&& \mbox{ if
$\beta-\Delta\alpha-\Delta\beta\leq\alpha\leq\beta$}\\
c_{\beta-\Delta\beta}-c_{\beta},&& \mbox{ if $\alpha>\beta$}
\end{array}\right.
 \end{equation}
In the first regime $\alpha<\beta-\Delta\alpha-\Delta\beta$, the
function $f_2$ is obviously negative ($\Delta\alpha> 0$), while in
the third regime $\alpha>\beta$, $f_2$ is constant and positive.
(Different types of typical behavior of $f_1$, $f_2$ and $f_3$ are
depicted in Fig. \ref{fig1} for illustration.)

Let us take a closer look at the second regime
$\beta-\Delta\alpha-\Delta\beta\leq\alpha\leq\beta$. First, solving
$f_2=c_{\beta-\Delta\beta}-c_{\alpha}=0$ for $\alpha$ gives
$\alpha_{\ast}=\beta-\Delta\beta$. Then, Eq. (\ref{ent_balance})
immediately implies that $\Delta\alpha=\Delta\beta$, and therefore
$f_1=f_3=0$, i.e. $\alpha_{\ast}$ is indeed the point where all
three functions simultaneously cross the $\alpha$ axis. It is
straightforward to see that $f_2$ is negative when
$\beta-\Delta\alpha-\Delta\beta<\alpha\leq\alpha_{\ast}$ and
positive when $\alpha_{\ast}<\alpha\leq\beta$.


Unlike $f_2$, it is not so easy to show when $f_1<0$. The main
reason is that $f_2$ is expressed in terms of three parameters at
most. $f_1$, however, involves all four parameters. Although only
three of them are free parameters, they are related by the implicit
equation (\ref{ent_balance}) and there is no simple analytic way to
express one of them, say $\Delta\alpha$, in terms of the others.
Thus, the negativity of $f_1$ cannot be demonstrated by the simple
substitution of $\Delta\alpha$ into Eq. (\ref{f1}). Therefore, I
will tackle the problem in a different way. I will show that the
first derivative of $f_1$, with respect to $\alpha$, is negative in
the whole interval $0\leq\alpha\leq (\pi/4-\Delta\alpha)$, i.e.
$f_1$ is strictly decreasing. This result will lead me to the
following conclusions: (a) The fact that the first derivative of the
continuous function $f_1$ does not change the sign is sufficient to
conclude that no other roots of $f_1=0$, except $\alpha_{\ast}$,
exist in that interval, i.e. $f_1$ crosses the $\alpha$-axis only at
$\alpha_{\ast}$ \cite{footnote_2}, (b) $f_1$ is positive below
$\alpha_{\ast}$ and negative above $\alpha_{\ast}$. Now, let us find
out the sign of the first derivative of $f_1$.
\begin{equation}\label{f1_firstder}
{df_1 \over
d\alpha}=s_{\alpha}c_{\beta}-s_{\alpha+\Delta\alpha}c_{\beta-\Delta\beta}
\left ( 1+{d\Delta\alpha \over d\alpha} \right ).
\end{equation}
From Eq. (\ref{ent_balance}) we obtain a relation for
$d\Delta\alpha/ d\alpha$. The differentiation of Eq.
(\ref{ent_balance}) in respect to $\alpha$ gives
\begin{equation}\label{1stder}
s_{2\alpha}\ln \tan \alpha=s_{2(\alpha+\Delta\alpha)}\ln
\tan(\alpha+\Delta\alpha)
 \left ( 1+{d\Delta\alpha \over
d\alpha} \right ).
\end{equation}
Now, let us substitute Eq. (\ref{1stder}) into Eq.
(\ref{f1_firstder});
\begin{equation}\label{f1_firstder_final}
{df_1 \over
d\alpha}=s_{\alpha}\left[c_{\beta}-c_{\beta-\Delta\beta}\frac{c_{\alpha}}{
c_{\alpha+\Delta\alpha}}\frac{\ln\tan\alpha}{\ln\tan(\alpha+\Delta\alpha)}\right].
\end{equation}
The second term in square brakets is a factor of three products.
This term is larger than the first term, $c_{\beta}$, because the
first factor, $c_{\beta-\Delta\beta}$, is larger than $c_{\beta}$,
while the other two factors are larger than $1$. That implies that
$d f_1/d \alpha<0$.

To summarize, we have shown that for $0\leq\alpha<\alpha_{\ast}$ the
function $f_2$ is negative, while for
$\alpha_{\ast}<\alpha\leq(\pi/4-\Delta\alpha)$ the function $f_1$ is
negative, and $\alpha_{\ast}$ is the only point where all three
inequalities of the majorization condition are satisfied. This value
corresponds to the situation of state swapping described in Sec.
\ref{intro}. Indeed, Eq. (\ref{ent_balance}) implies that
$\Delta\alpha=\Delta\beta$ for $\alpha=\alpha_{\ast}$, i.e., $\Delta
E=E(\psi_{\beta})-E(\phi_{\alpha})$.

Note, that we did not analyze here the sign of $f_3$ analytically.
The signs of $f_1$ and $f_2$ were sufficient to prove the main
result. The typical behavior of $f_3$ can be seen, though, from the
numerical simulations presented in Fig. \ref{fig1}, and will be
discussed in more detail in the next section.

\section{Entanglement transfer with catalysis}\label{catalysis}
Some transformations that are impossible under LOCC become possible
in the presence of a {\it catalytic} state \cite{catalysis}. In this
section I address the question of whether catalysis can help in our
case.

It was proved that catalysis can help only if the initial total
state $|\psi_{\beta}\ra_{AB}\otimes|\phi_{\alpha}\ra_{ab}$ and the
final total state
$|\psi_{\beta-\Delta\beta}\ra_{AB}\otimes|\phi_{\alpha+\Delta\alpha}\ra_{ab}$
are {\it incomparable} \cite{catalysis}. For a $4\times4$-level
system the necessary conditions for the possibility of catalytic
transformation are
\begin{equation}\label{cat_cond}
f_1\geq 0, ~~~~~~ f_2< 0, ~~~~~~ f_3\geq 0.
\end{equation}

From the results of the previous section it follows that the first
two conditions are not satisfied if $\alpha>\alpha_{\ast}$.

When $\alpha<\alpha_{\ast}$, however, the first two conditions in
Eq. (\ref{cat_cond}) are satisfied and the possibility of catalytic
transformation depends on the sign of $f_3$. As we see from Fig.
\ref{fig1}, $f_3$ can take positive values in some cases. The
analytic analysis of the sign of $f_3$ would be more difficult than
that of $f_1$ and $f_2$. I will combine numerical and analytical
techniques instead. As we can see from Fig. \ref{fig1}, $f_3$ takes
positive values at $\alpha<\alpha_{\ast}$ only if $\beta$ is larger
than a certain value. This critical $\beta_c$ corresponds to the
point where two roots of $f_3=0$ are degenerate [Fig.
\ref{fig1}(b)]. We notice that for $\beta=\beta_c$, the derivative
of $f_3$ in respect to $\alpha$ is zero at $\alpha=\alpha_{\ast}$.
We will use this fact to deduce the value of $\beta_c$;
\begin{equation}\label{f2_firstder}
{df_3 \over
d\alpha}=c_{\alpha}s_{\beta}-c_{\alpha+\Delta\alpha}s_{\beta-\Delta\beta}
\left ( 1+{d\Delta\alpha \over d\alpha} \right ).
\end{equation}
Substituting Eq. (\ref{1stder}) into Eq. (\ref{f2_firstder}) we
obtain
\begin{equation}
{df_3 \over
d\alpha}=c_{\alpha}\left(s_{\beta}-{s_{\alpha}s_{\beta-\Delta\beta}\over
s_{\alpha+\Delta\alpha}} {\ln\tan\alpha\over
\ln\tan(\alpha+\Delta\alpha)}\right).
\end{equation}
Therefore, at $\alpha=\alpha_{\ast}$ and $\beta=\beta_c$ we get
\begin{equation}
\left({s_{\beta_c}\over s_{\beta_c-\Delta\beta}}\right)^2 =
{\ln\tan(\beta_c-\Delta\beta)\over \ln\tan \beta_c}.
\end{equation}
For a given $\Delta\beta$ we solve this equation numerically. For
$\Delta\beta=0.01$ used previously in numerical examples (Fig.
\ref{fig1}) $\beta_c=0.490549$. Figure 2 shows $\beta_c$ as a
function of $\Delta\beta$. $\beta_c$ approaches $0.48557$ as
$\Delta\beta\rightarrow 0$.

\begin{figure} \epsfxsize=3.5truein
      \centerline{\epsffile{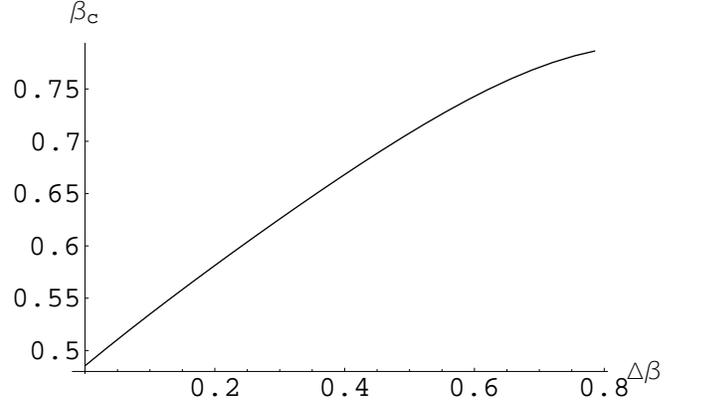}}
  \caption[]{$\beta_c$ as functions of $\Delta\beta$.}
    \label{fig2} \end{figure}

\begin{figure} \epsfxsize=3.5truein
      \centerline{\epsffile{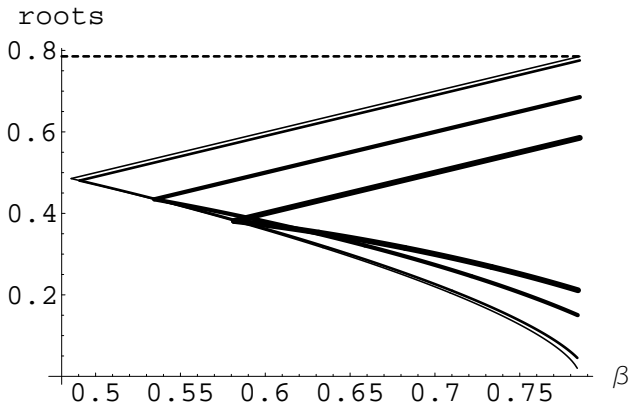}}
  \caption[]{Two roots of $f_3=0$ as functions of $\beta$ for $\Delta\beta=0.2; 0.1; 0.01; 0.001$
  (the thinner line corresponds to smaller $\Delta\beta$) and $\beta>\beta_c$. The
  upper line in each pair corresponds to $\alpha_{\ast}$. The lower line corresponds to the second
  root. All $\alpha$ that lie between these two roots are allowed in
  the presence of the catalytic state. The dashed line corresponds to $\alpha=\pi/4$.}
    \label{fig3} \end{figure}

Thus, only for $\beta>\beta_c$ the catalytic transformation is
possible. The range of $\alpha$, which allows this, is confined to
the interval between two roots of $f_3=0$. The first (larger) root,
as we have shown analytically, is $\alpha_{\ast}$. The second
(smaller) root can be obtained by solving simultaneous equations
$f_3=0$ and Eq. (\ref{ent_balance}) numerically. For example, in the
case presented in Fig. \ref{fig1}(c), the range of allowed $\alpha$
is $[0.3274, \pi/5-0.01]$. Figure \ref{fig3} presents both roots as
a function of $\beta$ for
 $\beta>\beta_c$ and four different values of $\Delta\beta$. We see that for a given $\Delta\beta$ the range of
 allowed $\alpha$ broadens towards larger $\beta$. The situation improves as $\Delta\beta$ decreases, and for very small
 $\Delta\beta$ the catalytic transformation becomes possible for
 essentially all values of $\alpha$ as $\beta$ becomes
 close to $\pi/4$ (i.e., $|\psi_{\beta}\ra$ is a nearly maximally
 entangled state).

\section{Probabilistic entanglement transfer}\label{prob_transfer}
\begin{figure} \epsfxsize=3.5truein
      \centerline{\epsffile{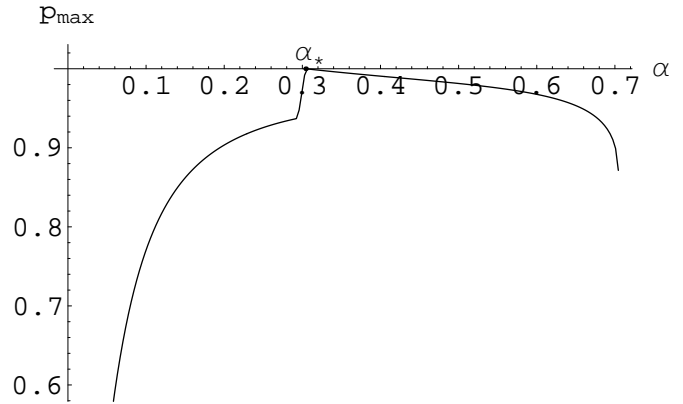}}
  \caption[]{$p_{max}$ vs $\alpha$ dependence for $\Delta\beta=0.01$ and $\beta=\pi/10$. $p_{max}$
  reaches $1$ at $\alpha=\alpha_{\ast}$.}
    \label{fig4} \end{figure}

So far we have seen that a reliable entanglement transfer is very
restricted. To complete our analysis it is worth mentioning how the
situation might be improved if we allow the transfer to be
accomplished with some probability of success less than 1. How close
to 1 can we get? To this end we use the extension of the
majorization condition to the probabilistic transformations of a
single copy \cite{lopopescu,vidal}, which in our case implies that
the maximum probability of a successful transformation is
\begin{eqnarray}
p_{max}=min\left\{\frac{1-\lambda_1}{1-\lambda_1'},\frac{1-\lambda_1-\lambda_2}{1-\lambda_1'-\lambda_2'},
\frac{\lambda_4}{\lambda_4'}\right\}.
\end{eqnarray}
As an example, let us consider the combination of parameters
described in Fig. \ref{fig1}(a). It can be easily checked that
$p_{max}={1-\lambda_1 \over 1-\lambda_1'}$ for
$\alpha>\alpha_{\ast}$, while
$p_{max}=\frac{1-\lambda_1-\lambda_2}{1-\lambda_1'-\lambda_2'}$ for
$\alpha<\alpha_{\ast}$. The $p_{max}$ vs $\alpha$ dependence is
depicted in Fig. \ref{fig4}.
The probability of failure that must be accepted in order to allow
the entanglement transfer is finite and increases with the distance
between $\alpha$ and $\alpha_{\ast}$.

\section{Asymptotic case}\label{asymptotic}
Not surprisingly, all limitations described in the previous sections
disappear in the asymptotic limit. We assume that Alice and Bob
possess $n\rightarrow\infty$ copies of $|\psi_{\beta}\ra$ and Alex
and Barbara possess $n$ copies of $|\phi_{\alpha}\ra$.

Clearly, Alice and Bob are able to obtain $n$ copies of
$|\psi_{\beta-\Delta\beta}\ra$ without any help from Alex and
Barbara simply by using the asymptotic entanglement
concentration(distillation) method \cite{proc_meth}: first they
concentrate $|\psi_{\beta}\ra^{\otimes n}$ into $n H(c_{\beta}^2)$
singlets and then distill them into
$|\psi_{\beta-\Delta\beta}\ra^{\otimes
nH(c_{\beta}^2)/H(c_{\beta-\Delta\beta}^2)}$, thus apart from
required $n$ copies they obtain
$n[H(c_{\beta}^2)/H(c_{\beta-\Delta\beta}^2)-1]$ additional copies
of $|\psi_{\beta-\Delta\beta}\ra$.

Alex and Barbara are able now to absorb the entanglement of these
additional copies into their states. First, they distill (or
concentrate, depending on the value of $\alpha$)
$|\phi_{\alpha}\ra^{\otimes n}$ into
$|\phi_{\beta-\Delta\beta}\ra^{\otimes
nH(c_{\alpha}^2)/H(c_{\beta-\Delta\beta}^2)}$. Now, acting
collectively on these copies together with
$n[H(c_{\beta}^2)/H(c_{\beta-\Delta\beta}^2)-1]$ copies of
$|\psi_{\beta-\Delta\beta}\ra$, they  concentrate them into $n$
copies of $|\phi_{\alpha+\Delta\alpha}\ra$, where $\Delta\alpha$
satisfies Eq. (\ref{ent_balance}). Thus, we obtain
$\left(|\psi_{\beta-\Delta\beta}\ra_{AB}\otimes|\phi_{\alpha+\Delta\alpha}\ra_{ab}\right)^{\otimes
n}$ as desired.

This procedure is a clear demonstration that there are no
restrictions on the redistribution of entanglement in the asymptotic
limit.

\section{Conclusion}\label{concl}
In this paper I have analyzed the question of reliable entanglement
transfer between two bipartite pure states, which I call the donor
and the acceptor states. The case of a disentangled acceptor state
was considered for systems of any dimensionality. It was shown that
no partial entanglement can be transferred if the Schmidt number of
the donor state does not change.

In the case of qubit states it has been shown that the amount of
entanglement allowed to be transferred reliably to an entangled
acceptor state is very restricted. Without the presence of a
catalytic state the transfer is possible only when the entanglement
of the acceptor state $E(\phi)$ is smaller than the entanglement of
the donor state $E(\psi)$. The amount that can be transferred is
just the difference between the two, $\Delta E=E(\psi)-E(\phi)$. The
task is accomplished by swapping the states, which can be done
locally.
In all other cases, 
the transfer is impossible.

In the presence of a catalytic state the above restrictions are
relaxed to a certain degree. Transfer might be possible subject to
the following conditions. The first condition is $E(\phi)\leq
E(\psi)-\Delta E$. Note that this essentially implies that no
entanglement can be transferred to an acceptor state that is more
entangled than the donor state. The second condition is, for a given
$\Delta E$ the entanglement of the donor state has to be larger than
a certain threshold $E(\psi_{\beta_c })=f(\Delta E)$. The third
condition is, $E(\phi)$ has to fall into a certain range, which
broadens as $E(\psi)$ increases. As $E(\psi)$ tends to maximum and
$\Delta E$ is small, the catalytic transformation becomes possible
for all values of $E(\phi)$. Thus, using catalysis it is always
possible to ``chop" a small piece of entanglement from a maximally
entangled donor state and transfer it to an acceptor state
(providing the acceptor state has ``room" for this amount of
entanglement, of course). Alternatively, the entanglement of a
maximally entangled donor state can be transferred in full.

The above restrictions were derived under the requirement of a
reliable transfer. The possibility of a probabilistic transfer also
has been discussed, and it was shown that the probability of a
successful transfer cannot be made arbitrarily close to 1.
Reliability can also be sacrificed by allowing part of the
transferred entanglement to be lost. A preliminary analysis shows
that such losses are not negligible.

Entanglement transfer, addressed in this paper, should be compared
with the broadcasting of entanglement \cite{ent_broadcast} and
entanglement splitting \cite{ent_split}. To say that entanglement
has been (partially) broadcasted is to say that two less-entangled
states have been obtained from one more-entangled state by local
operations. To say that entanglement has been split is to say that
the entanglement of a pure state has been split into two branches,
i.e., the second party had ``shared" her entanglement with a third
party, so they are both now entangled with the first party. There
are two main differences between my approach and those of Refs.
\cite{ent_broadcast} and \cite{ent_split}. First, I require that the
resulting states remain pure. In Ref. \cite{ent_broadcast} this
requirement was relaxed, and the separability criterion for mixed
states was used to analyze the entanglement of the resulting states.
In fact, the results of Sec. \ref{product_acceptor} of this paper
imply that entanglement cannot be broadcasted to pure states. In
Ref. \cite{ent_split} the requirement of purity obviously cannot be
applied to a single branch. Second, entanglement is broadcasted to
initially disentangled particles (nonentangled acceptor state),
whereas entanglement transfer, analyzed here, takes place when the
acceptor-particles are entangled. In this sense, entanglement
transfer is more general.

I believe that the results of this paper have shed more light on the
nature of entanglement of a finite number of pure states. It will be
interesting to generalize this argument to a quantum system of
higher dimensionality. There is also a potential unexplored relation
between entanglement transfer and the broadcasting of entanglement
and entanglement splitting.

\begin{acknowledgments}
I am grateful to Roger Colbeck for his help. This work was funded by
the U.K. Engineering and Physical Sciences Research Council.
\end{acknowledgments}


\begin{thebibliography}{30}
\bibitem{footnote_1} Here and in the following I will use the notation $c_x\equiv\cos
x$, $s_x\equiv\sin x$.
\bibitem{footnote_2} Although, this statement is not proven mathematically
rigorously, its validity seems obvious.
\end{thebibliography}
\end{document}